\documentclass[aps,prb,twocolumn]{revtex4}
\usepackage{makeidx}
\usepackage{amsmath,amssymb,amsfonts,amsthm}
\usepackage{graphicx,bm}

\begin{document}

\title{Effects of aging and annealing on the polar and antiferrodistortive
components of the antiferroelectric transition in PbZr$_{1-x}$Ti$_{x}$O$_{3}$%
}
\date{}
\author{F. Cordero,$^{1}$ F. Trequattrini,$^{2}$ F. Craciun,$^{1}$ C. Galassi%
$^{3}$}
\affiliation{$^1$ CNR-ISC, Istituto dei Sistemi Complessi, Area della Ricerca di Roma -
Tor Vergata,\\
Via del Fosso del Cavaliere 100, I-00133 Roma, Italy}
\affiliation{$^{2}$ Dipartimento di Fisica, Universit\`{a} di Roma \textquotedblleft La
Sapienza\textquotedblright , P.le A. Moro 2, I-00185 Roma, Italy}
\affiliation{$^{3}$ CNR-ISTEC, Istituto di Scienza e Tecnologia dei Materiali Ceramici,
Via Granarolo 64, I-48018 Faenza, Italy}

\begin{abstract}
The antipolar and antiferrodistortive (AFD) components of the
antiferroelectric (AFE) transition in PbZr$_{1-x}$Ti$_{x}$O$_{3}$ ($x\leq
0.054$) can occur separately and with different kinetics, depending on the
sample history, and are accompanied by elastic softening and stiffening,
respectively. Together with the softening that accompanies octahedral
tilting in the fraction of phase that is not yet transformed into AFE, they
give rise to a variety of shapes of the curves of the elastic compliance
versus temperature. All such anomalies found in samples with $x=0.046$ and
0.054, in addition to those already studied at $x=0.050$, can be fitted
consistently with a phenomenological model based on the simple hypothesis
that each of the polar and AFD transitions produces a step in the elastic
modulus, whose position in temperature and width reflect the progress of
each transition. The slowing of the kinetics of the transformations is
correlated with the formation of defect structures during aging in the
ferroelectric (FE) or AFE state, which are also responsible for a
progressive softening of the lattice with time and thermal cycling, until
annealing at high temperature recovers the initial conditions.
\end{abstract}

\maketitle


\section{Introduction}

The AFE/FE transition in PbZrO$_{3}$ based perovskites involves large volume
changes and therefore is widely studied in view of applications for
actuators,\cite{Uch98} but also other applications are envisaged, such as
high energy storage capacitors or cooling devices.\cite{HZK14} The large
strains involved in the transition may be partly responsible for the slow
kinetics of the FE to AFE transition driven by cooling, and hence large
thermal hysteresis, and may even cause lattice damage ranging from broken
bonds\cite{PP99,PP07} to cracks.\cite{ZZZ04,Lou09b} The slow kinetics of
AFE/FE transitions has been the object of experimental and
theoretical/phenomenological modeling, including a series of articles by
Ishchuk and coworkers,\cite{IS11} who propose that short range cation
migration is involved, and is made possible at room temperature by the large
stress at the AFE/FE interfaces.\cite{ISS06} The physics of the AFE/FE
transition can be particularly rich, and these authors even suggest that the
relaxor transition can be explained in terms of coexisting FE and AFE phases.%
\cite{Ish01d}

An additional phenomenon often related to the AFE transitions is the
existence of incommensurate structural modulations. Local structural probes
like electron diffraction reveal that the AFE transition may be preceded by
incommensurate modulations of the cation displacements, and similar
coexistence of incommensurate modulated AFE structures with the paraelectric
or FE phase have been observed in Pb(Sc$_{1/2}$Ta$_{1/2}$)O$_{3}$,\cite{BB90}
and other perovskites, including variously doped\cite{RWR11} but also undoped%
\cite{WK01} PbZr$_{1-x}$Ti$_{x}$O$_{3}$ (PZT).

A new type of information came from the combination of elastic, dielectric
and diffraction experiments on PZT,\cite{150} that showed how the polar and
antiferrodistortive (AFD) degrees of freedom can act almost independently of
each other and with very different kinetics in the AFE/FE transition. In
fact, the transition from the rhombohedral (R) FE phase to the orthorhombic
(O) AFE phase involves both cation displacements and AFD tilting of the O
octahedra according to the $a^{-}a^{-}c^{0}$ pattern, in Glazer's notation.%
\cite{Gla72} While the FE and AFE cation modes are easily detectable by
diffraction and produce large effects in the dielectric response, the
transitions involving AFD modes produce clear steps in the elastic moduli,
which are also sensitive to the polar transitions. We showed that in PZT
with 5\% Ti the R-FE/O-AFE transition involves the concomitant condensation
or ordering of both AFD and polar modes in quasistatic experiments, but the
AFE component could be much slower or even absent when cooling at rates
above 0.5~K/min.\cite{150} The curves of the Young's modulus as a function
of temperature were found to reflect the progress of the various polar and
AFD transitions, and exhibited a variety of shapes, depending on the sample
history and temperature rate.

Here we extend the experiments to compositions of 4.6\% and 5.4\% Ti and
find that it is also possible that the AFD transition is slower than the AFE
one. Similarly to the previous work,\cite{150} to which we refer for the
region of interest of the phase diagram of PZT, the variety of shapes of
elastic anomalies found under different conditions can be fitted setting the
compliance as the volume average of the compliances of the four possible
phases coexisting in that region of the phase diagram, counting the $%
a^{-}a^{-}c^{0}$ tilt pattern and the AFE cation ordering as separate
phases, and including the tilted R-FE phase $R3c$. Here we added the option
of imposing that the O tilt pattern sets in only in the AFE phase. The slow
kinetics of the transformations are correlated with the formation of defect
structures during aging in the FE or AFE state, which are also responsible
for a progressive softening of the lattice with time and thermal cycling.

It may be interesting to note that in multiferroic BiFeO$_3$ there is a
nanoscale alternation of R-FE $R3c$ and O-AFE domains with the same tilt
patterns as in PZT (but with different AFE pattern of the Bi atoms),\cite%
{WSM13} and a modulated superstructure involving tilting and AFE
displacements of Ti is found in EuTiO$_3$, another multiferroic perovskite.%
\cite{KTB13} Likely, understanding the mechanisms of interaction between the
tilt and polar modes in AFE PZT provides insight into those operating in the
multiferroic perovskites and viceversa.

\section{Experimental}

Ceramic samples of PbZr$_{1-x}$Ti$_{x}$O$_{3}$ (abbreviated as PZT100$x$)
with $x=0.046$ and 0.054 were prepared in the same manner as previous series
of samples.\cite{127,145} The oxide powders were calcined at 800~$^{\circ }$%
C for 4 hours, pressed into bars, sintered at 1250~$^{\circ }$C for 2~h in
crucibles packed with PbZrO$_{3}$\ + 5wt\% excess ZrO$_{2}$\ to prevent PbO
loss. The powder X-ray diffraction did not reveal any trace of impurity
phases. The densities were about 95\% of the theoretical values and the
grains were large, with sizes of $5-20$~$\mu $m. The bars were $4~$cm long
and cut into slices $0.6$~mm thick. In order to measure the dynamic Young's
modulus $E\left( \omega \right) =$ $E^{\prime }-iE^{\prime \prime }$, the
sample was suspended on thin thermocouple wires in vacuum and
electrostatically excited on its first odd flexural modes with an electrode
close to its surface, partially made conducting with Ag paint. Sample and
electrode also constitute a capacitor inserted in an oscillator whose
high-frequency of $\sim 150$~kHz is modulated by the sample vibration and
can be demodulated to produce a signal proportional to the sample vibration.%
\cite{135} Below we present the reciprocal of the Young's modulus, the
compliance $s\left( \omega ,T\right) =s^{\prime }-is^{\prime \prime }$,
which is the mechanical analogue of the dielectric susceptibility, measured
on the fundamental mode with $\omega /2\pi $ $\sim 1.6$~kHz. Since $\omega
\propto \sqrt{E^{\prime }}$,\cite{NB72} the temperature variation of $s$ is
given by $s\left( T\right) /s_{0}\simeq $ $\omega _{0}^{2}/\omega ^{2}\left(
T\right) $, where $\omega _{0}$ is chosen so that $s_{0}$ is the compliance
in the paraelectric phase. The elastic losses are presented as $s^{\prime
\prime }/s^{\prime }=Q^{-1}$, where the mechanical $Q$ of the sample was
measured from the width of the resonance curve or from the decay of the free
oscillations after switching off the excitation.

The dielectric permittivity $\varepsilon \left( \omega ,T\right)
=\varepsilon ^{\prime }-i\varepsilon ^{\prime \prime }$ was measured with a
HP 4284A LCR meter with a four-wire probe and an electric field of 0.5 V/mm,
between 0.2 and 200~kHz on disc samples with 12~mm diameter and 0.7~mm
thick. Temperature was controlled with a modified Linkam HFS600E-PB4 stage.

\section{Results and Discussion}

The same notation as in the previous study on PZT5 will be adopted. We do
not attribute space groups to the intermediate phases undergoing the partial
AFE or AFD transformations for two reasons. One is that the dielectric and
elastic susceptibilities do not provide information on the cell structure,
but only on the type of the transformation. The other reason is that, as
discussed in Sect. \ref{stiff}, even the starting R-FE phase for $x\left(%
\text{Ti}\right) <$ 0.06 has an average $R3m$ structure, but the octahedra
should be already tilted below $T_{\mathrm{C}}$, though without long range
order.\cite{145,148} Therefore, already the description of the starting FE
phase might require two symmetry groups: one for the average structure
determined by neutron and X-ray diffraction, and lower symmetries for the
local structure. The structure within the broad temperature region of the
transformation between the FE $R3m$ and AFE $Pbam$ would require an even
more complex description, which we simply denote as $\sim Pbam$.

\begin{table*}[htb]
\caption{Transition temperatures with meaning and phases that they separate.
Superscript h/c = heating/cooling; L/SRO = long/short range order. It is
assumed that for $<6$\% Ti it is $T_{\mathrm{IT}}=$ $T_{\mathrm{C}}$ and
therefore the $R3m$ phase is tilted with SRO.\protect\cite{145,148}}
\label{Ttable}%
\begin{ruledtabular}
\begin{tabular}{lll}
$T_{\mathrm{C}}$ & Curie temperature (FE $\leftrightarrow $ PE)& $R3m$ $\leftrightarrow $ $Pm3m$\\
$T_{\mathrm{T}}$ & tilting (LRO $a^{-}a^{-}a^{-}$) & $R3c$ $%
\leftrightarrow $ $R3m$ (untilted or SRO tilted) \\
$T_{\mathrm{AF}}$ & AFE $\leftrightarrow $ FE (cation displacements) & $%
\sim Pbam$ $\leftrightarrow $ $R3m$ \\
$T_{\mathrm{OT}}$ & orthorhombic tilt (LRO $a^{-}a^{-}c^{0}$) & $\sim Pbam$ $%
\leftrightarrow $ $R3m$ \\
$T_{\mathrm{IT}}$ & intermediate tilt = SRO tilting & $R3m$ SRO tilted $%
\leftrightarrow $ $R3m$ untilted%
\end{tabular}
\end{ruledtabular}
\end{table*}

The nomenclature of the various transitions is detailed in Table I, and the
following abbreviations will be used: AF indicates the polar component of
the AFE/FE transition; OT (= octahedral tilting $a^{-}a^{-}c^{0}$) the AFD
component; T the tilting transition within the FE-R phase between untilted $%
R3m$ and $a^{-}a^{-}a^{-}$ tilted $R3c$. The transition temperatures will be
identified with the centers of the steps of $s^{\prime} \left( T\right)$ and
called $T_k^{h/c}$ where $k=$ AF, OT, T and $h/c$ specify heating/cooling.

\subsection{Characters of the transformations}

Both the FE/PE transition at $T_{\mathrm{C}}$ and the AFD at $T_{\mathrm{T}}$
involve a single order parameter (OP) and can in principle be of the second
order. Indeed, they exhibit little hysteresis between heating and cooling.
Instead, the transition between the FE $R3m$ or $R3c$ and the AFE $Pbam$
phases must be first order and involves two separate OPs. The polar mode
passes from predominant displacements of Ti along $\left( 111\right) $
(pseudocubic setting) in the FE phase to predominant displacements of Pb
along $\left\langle 1\overline{1}0\right\rangle $ with wave vector $\left(
\frac{1}{4}\frac{1}{4}0\right) $ in the AFE phase, without group-subgroup
relationship between the two phases.\cite{FH84,GRD93} An additional AFD
order parameter is responsible for the $a^{-}a^{-}c^{0}$ tilt pattern. In
pure PbZrO$_3$ the transition is ideally from the cubic phase, and the
temperature dependence of the two OPs has been described within the Landau
formalism in terms of a primary polar OP and a secondary AFD OP bilinearly
coupled with the former and nearly following it.\cite{FIT03}

The situation with 5\% Ti doping has been shown to be quite different,\cite%
{150} not only because the parent phase is R-FE instead of cubic PE, but
also because the AFD component of the transition can occur while the
antipolar one is frozen or hindered, suggesting exchanged roles of the two
OPs. Yet, at the two compositions studied here, very close to 5\% Ti, also
the reverse is true: the AFE OP may advance the AFD one. Evidently, at these
compositions near the morphotropic phase boundary with the FE phase, the
free energy is rather flat between the FE and AFE minima, and also with
respect to various tilt patterns, so that these minima can be easily
perturbed by slight changes in Zr/Ti distributions, extrinsic defects and
possibly even by twin and domain walls. This results in a readiness to
change the balance and kinetics of the two OPs upon small changes in the
sample preparation and history, and renders the description of the evolution
of the AFE phase transition much less transparent than for pure or lightly
doped PbZrO$_{3}$.\cite{FIT03}

As mentioned in the Introduction and in the previous
study,\cite{150} observations of exceptionally large thermal
hystereses and slow kinetics of the AFE transition in undoped and
variously doped PZT have been done with various techniques. A large
thermal hysteresis and coexistence of $R3m$, $R3c $ and $Pbam$
phases result from dielectric and pyroelectric measurements on PZT
with $x\left( \text{Ti}\right) =0.05$\cite{HU95} and with Raman
scattering in Nb-doped PZT with $x\leq 0.03$,\cite{KOD99} while TEM
observations\cite{VLD96,WK02} provide additional information on the
local octahedral tilting, and are consistent with the idea that the
$R3m$ phase at low $x$ is already tilted, though with a disordered
or incommensurate pattern. The nucleation of the AFE phase has been
followed also directly with a polarized microscope,\cite{DRS96}
finding that its athermal character in pure PbZrO$_{3}$, namely
independent of thermal fluctuations and hence unchanged under
stationary conditions, is changed by the Ti substitution into
strongly isothermal, where the evolution can slowly proceed at
constant temperature.\cite{DRS96} Codoping with La and Li on the Pb
site adds new possible mechanisms of aging on longer time scales,
related to the cation distribution, and the resulting changes at the
atomic level have been followed through the evolution of the
profiles of the x-ray diffraction peaks.\cite{ISS06}

No attempt at quantitative interpretations of the above and similar
observations are known to us, and no hint to a separation of the
kinetics of the AFE and AFD OPs. It turns out that the almost ideal
case of athermal transition in pure PbZrO$_{3}$, amenable to a full
interpretation within the Landau scheme, becomes considerably more
complex with the substitution of Ti and other dopants, and we will
limit our interpretation to the phenomenological level.

\subsection{Elastic compliance during cycling, aging and annealing}

\begin{figure}[tbh]
\includegraphics[width=8.5 cm]{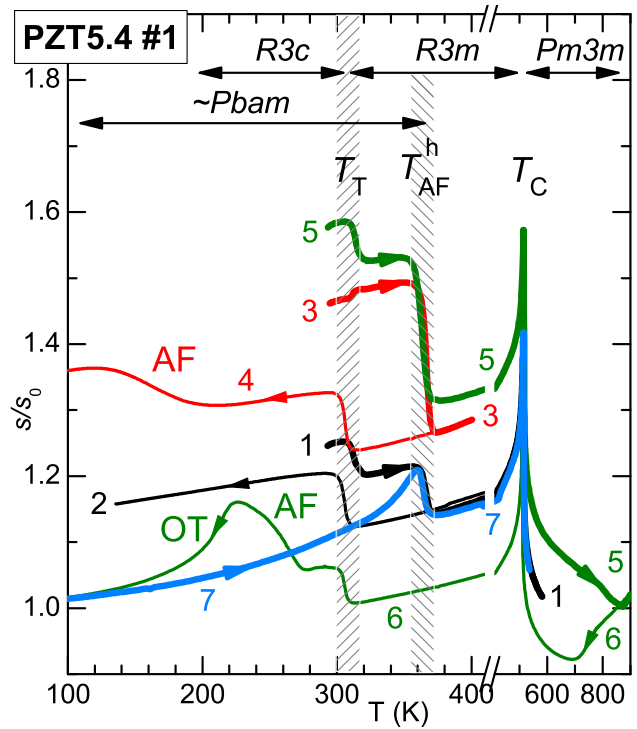}
\caption{Sequence of $s^{\prime }\left( T\right) $ curves of a sample of
PZT5.4 with the same normalization factor $s_{0}$ corresponding to 4.2~kHz.
The numbers indicate the sequence of runs; 3 was 47 days after 1, and 5
after additional 7 days. AF and OT indicate the polar and AFD (tilt)
components of the transition to the AFE-O $Pbam$ state.}
\label{ti5.4n1}
\end{figure}

Figure \ref{ti5.4n1} presents a sequence of $s^{\prime }\left( T\right) $
curves measured on PZT5.4 \#1 in order to demonstrate the main features of
the AFE and tilt transitions and the influence of room temperature aging and
high-temperature annealing on their kinetics. Only few of the measuring
cycles are shown, in order to avoid excessive crowding of the figure.
Temperature was changed at rates of $\geq 1$~K/min.

The shaded stripes mark the temperature ranges of the tilt instability in
the R phase at $T_{\mathrm{T}}$, and the AFE to FE transition during
heating, while the OT and AF orderings during cooling may appear at very
different temperatures or may even be absent. During the first heating
(curve 1), sharp stiffening is found both at $T_{\mathrm{T}}^{h}=314$~K and $%
T_{\mathrm{AF}}^{h}=365$~K; this means that the AFE phase was only partially
formed, while the rest remained $R3c$ and became $R3m$ at $T_{\mathrm{T}%
}^{h} $. The measurement was extended to 600~K, well above $T_{\mathrm{C}%
}=514$~K and enough to erase any memory of the AFE domains. In fact, the
subsequent cooling (2) exhibits only the softening at $T_{\mathrm{T}}^{c}$
and no trace of further transitions down to 140~K. This is the usual state
of our samples with $x\simeq 0.05$, with the AF/OT transitions very sluggish
during cooling. Curve 3 was measured after 47 days, during which some
measurements were done without exceeding $T_{\mathrm{C}}$. Apparently there
was sufficient time for almost complete transformation into AFE/OT phase. In
fact, there is only a hint of transformation at $T_{\mathrm{T}}$, and the
amplitude of the step at $T_{\mathrm{AF}}^{h}$ is much enhanced, but this
will be discussed more in depth in reference to Fig. \ref{plotP45}. Part of
the enhancement of the AF step, however, should be due to the fact that
aging rises also the overall compliance and particularly that of the AF
phase. This appears from the continuous increase with time of $s^{\prime
}\left( 400~\text{K}\right) $, outside the stability region of the AFE
phase. Curve 3 was interrupted at 400~K, apparently before completely
erasing the imprints of the AFE domains, because during cooling (4) a broad
AF transition was observed at $T_{\mathrm{AF}}^{c}=165$~K. Further cycling
(not shown)\ demonstrated that the transition to the AFE state was not
complete, and also after additional 6 days curve 5 presented stiffening at $%
T_{\mathrm{T}}^{h}$, in addition to that at $T_{\mathrm{AF}}^{h}$. Notice
that the step at $T_{\mathrm{T}}^{h}$ in curve 5 is due only to a minority
fraction of R-FE phase, but its magnitude is comparable with that of curve
2, where 100\% of the sample undergoes the tilt instability. This is again
evidence of a general enhancement of the compliance in all phases during
room temperature aging, a fact that we tentatively explain in terms of
formation of extended defects that soften the lattice. Certainly a
progressive softening with aging is opposite to the usual behavior of glassy
systems, where aging causes a decrease of the susceptibilities.

Curve 5 was extended to 900~K, and the subsequent cooling, 6, presents the
general effect of high temperature annealing: a general restiffening to the
original values and an acceleration of the kinetics of the transitions. Now
both the AF softening and OT stiffening are observed during cooling, and the
absence of stiffening at $T_{\mathrm{T}}$ during the subsequent heating 7
demonstrates that the transitions were complete.

\subsection{Extended defects}

A microscopic model of the defect structures that determine the kinetics of
the AFE transition in PZT and are probably involved in the progressive
softening is not available yet, but it may be related to the observations of
Pokharel and Pandey\cite{PP99} on the AFE transition in Pb$_{1-x}$Ba$_{x}$ZrO%
$_{3}$. Also in that material, the AFE/FE transformation presents huge
thermal hysteresis and important ageing effects over months, which have been
attributed to the lattice damage generated by the large transformation
strain. In particular, charged defects like dangling bonds would stabilize
the FE domains over the AFE ones, making the kinetics slow. Irreversible
damage like cracking and chemical decomposition certainly occurs under
repeated polarization cycles and is at the origin of fatigue in capacitors
and actuators, but some degree of damage may be expected to accompany even a
single FE/AFE transition driven by the change of temperature, if Acoustic
Emission is detected.\cite{DMC10} Yet, there are also TEM studies indicating
that there are no dislocations in correspondence with AFE/FE boundaries, and
the lattice framework is continuous.\cite{ISS06} It has been proposed that
the resulting large strain between small volume AFE and large volume FE
domains promotes the short range diffusion of cations with large difference
in size.\cite{IS02b,ISS06} Such cation exchanges require the overcoming of
barriers of several eV, and therefore are practically impossible at room
temperature in a homogeneous material, but the situation may be different in
the highly out-of-equilibrium process of nucleation of AFE phase from the FE
one. Yet, it is doubtful that such chemical inhomogeneities can be erased by
simple thermal activation during short annealings at 900~K, as in the
present measurements; in fact, in the paraelectric phase there cannot be
strong fields that cause out-of-equilibrium cation diffusion, and the
cations should diffuse at the much higher temperatures where sintering
occurs. In addition, cation diffusion has been proposed in PZT codoped with
La$^{3+}$ and Li$^{1+}$ on the Pb$^{2+}$ site, creating additional cation
species widely differing in size and charge and therefore prone to
rearrange, while in our case only Zr$^{4+}$ and Ti$^{4+}$ might undergo
rearrangements.

The most obvious candidates to explain our observations of irreversible
softening and recover after moderate annealing are charged V$_{\text{O}}$
that migrate in the internal electric fields of the FE and AFE domains and
form stable structures, for example rows or even planes. This would occur
mainly at twin and domain walls,\cite{TXL97,LSB05} where the electric and
stress fields and their gradients are maximal and the interactions with the
electric charges and dipoles and elastic quadrupoles of the defects are
strongest. Such extended defects would in turn pin the walls and act as
template for the formation of the AFE phase during subsequent
transformations. That these defect structures conform to the AFE phase,
during its slow formation over several days, is confirmed by the fact that
the transition to the AFE phase remains easier after a brief excursion into
the FE phase (curve 4 in Fig. \ref{ti5.4n1}). On the other hand, these
extended defects must be the cause of the progressive lattice softening
during aging, and are completely dissolved by annealing up to 900~K. A brief
excursion to 900~K in vacuum cannot heal cracks or recover a decomposed
perovskite phase, and for this reason we believe that we are dealing with
reversible defects like clusters of preexisting V$_{\text{O}}$, rather than
additional loss of PbO, cracks or cation diffusion. In order to verify that
no appreciable loss of PbO or oxygen occurs during these cyclings, which are
made at $2.5-4$~K/min when exceeding the Curie temperature, we repeated them
in a thermobalance and found no mass loss above the background value.

\begin{figure}[hbt]
\includegraphics[width=8.5 cm]{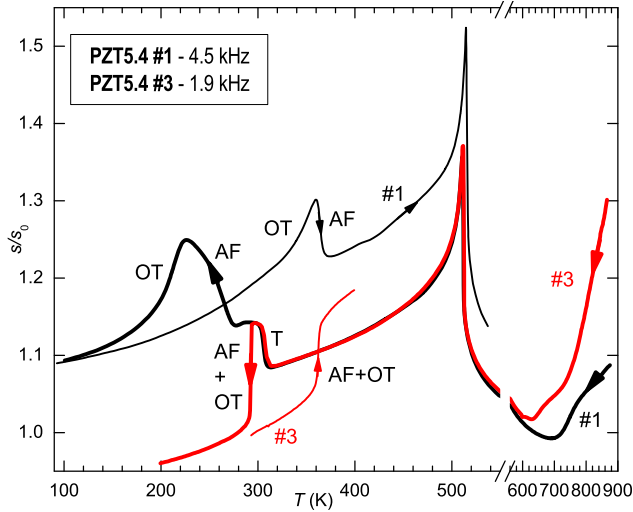}
\caption{Elastic compliance of two samples of PZT5.4 measured during cooling
from 900~K (thick lines), and subsequent heating (thin lines).}
\label{ti5.4n1n3}
\end{figure}

These observations are in agreement with the work of Zhou \textit{et al.,}%
\cite{ZZZ04} where partial recovery from fatigue after $10^{8}$ strain
hysteresis loops is obtained by annealing 1~h at 500~$^{\circ }$C. Also in
that case it is concluded that migration and redistribution of charged
species such as O vacancies must be involved, rather than healing of cracks.
What is noticeable in the present experiments is that huge effects like
softening of a factor of three do not occur after $10^{8}$ strain hysteresis
loops, but after aging for few weeks, during which few thermal cycles are
made.

Note that a quick anneal to 900~K recovers the original state of a sample
but does not not completely erase differences from sample to sample. This is
shown in Figure \ref{ti5.4n1n3}, comparing the $s^{\prime }\left( T\right) $
curves of two samples with $x=0.054$, after cycling up to 900~K in vacuum.
The curves of sample \#1 are the same as curves 6 and 7 of Fig. \ref{ti5.4n1}%
, while the curves from sample \#3 clearly display AF and OT transitions
that are very sharp and occur at exactly the same temperature both during
heating and cooling. Note also the perfect reproducibility of the curves
after annealing from the FE transition down to the step at $T_{\mathrm{T}%
}^{c}$. While Figs. \ref{ti5.4n1} and \ref{ti5.4n1n3} present cases in which
the AF and OT components are sharp and coincident (PZT5.4 \#3) or OT is
slower and broader than AF (PZT5.4 \#1), the third case of a slower AF
component was found in samples with $5\%$~Ti.\cite{150}

\subsection{Dielectric permittivity}

In the previous work on PZT5 it was demonstrated by comparison with X-ray
diffraction and dielectric spectroscopy how the softening during cooling is
associated with the antipolar ordering, while the setting of the $%
a^{-}a^{-}c^{0}$ tilt pattern causes a stiffening. The latter could be
observed as sharp steplike anomalies even when no sign of antipolar order
was observed, at cooling rates $>0.5$~K/min. The present samples with Ti
compositions slightly above and below 5\%Ti display a nearly opposite
behavior: during cooling at $\geq 1$~K/min the AF component is observed and
starts before the OT component. We do not have an explanation for the
different behaviors in the two sets of samples, prepared under very similar
conditions, and checked with dielectric spectroscopy that indeed in the
present samples the AF order is established also during relatively fast
cooling.

\begin{figure}[hbt]
\includegraphics[width=8.5 cm]{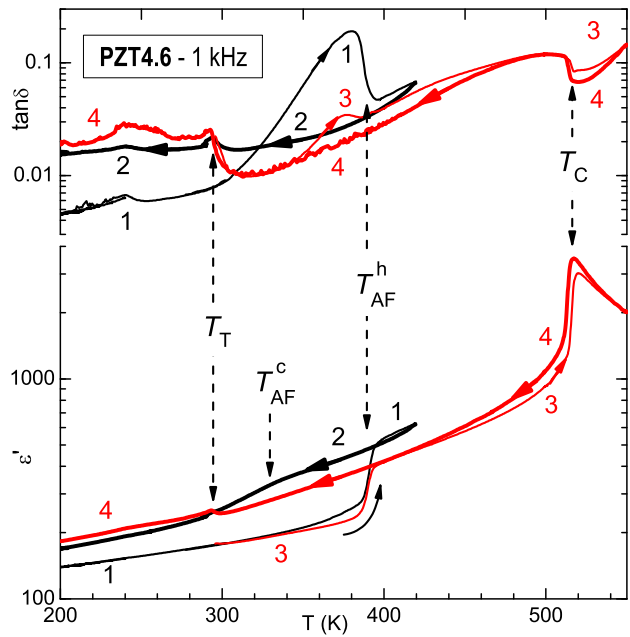}
\caption{{Dielectric permittivity of PZT4.6 measured at $\pm 1.5~$K/min}.
The numbers indicate the sequence of temperature runs.}
\label{diel4.6}
\end{figure}

Figure \ref{diel4.6} contains the dielectric spectra of PZT4.6 measured at
rates of 1.5~K/min during two cycles, in the virgin state and in the next
day. Initially the sample was in the AF state and the heating curve 1
presents a clear step at $T_{\mathrm{AF}}^{h}$; the rate was inverted
already at 420~K, and the opposite step is observed at $T_{\mathrm{AF}%
}^{c}\simeq 330$~K (curve 2). Though smaller and broader than the step at $%
T_{\mathrm{AF}}^{h}$, it clearly demonstrates that the AF transition indeed
occurs also during cooling at 1.5~K/min in this sample. The subsequent
heating curve 3 closely follows the previous one, 1, and was extended above $%
T_{\mathrm{C}}$, so erasing any memory of the AF state, and in fact no trace
of AF transition is found during the cooling 4. The anomalies in both real
part and losses at $T_{\mathrm{T}}$ are due to the sharp octahedral tilting
transition in the FE-R phase, while, as in the previous study, no signature
is associated with OT tilting. The dielectric and anelastic measurements are
therefore in agreement regarding the various transformation kinetics.

\subsection{Stiffening from the O tilt order\label{stiff}}

The fact that stiffening rather than softening is observed at the OT
transition deserves a comment. In fact, according to the established
opinion, the R-FE phase of PZT is untilted above $T_{\mathrm{T}}$ and
therefore tilting to the OT pattern should be accompanied by steplike
softening. This derives from the Landau theory of phase transitions with
linear-quadratic coupling $\lambda \varepsilon q^{2}$ between strain $%
\varepsilon $ and order parameter, the tilt angles $q$ of the octahedra, in
this case. The resulting elastic anomaly is a negative step $\propto
-\lambda ^{2}/c$, where $c$ is the elastic constant of $\varepsilon $.\cite%
{Reh73,KHC09} Other two cases have been reported where stiffening instead of
the expected softening is observed at a tilt-type transition: Ca$_{1-x}$Sr$%
_{x}$TiO$_{3}$ and SrZrO$_{3}$ passing from a tetragonal tilted high
temperature phase to the orthorhombic $a^{-}a^{-}c^{+}$ tilted phase. The
stiffening has been tentatively explained as due to a mutual blocking of the
two order parameters having different irreducible representations, namely
anti-phase tilting along $a$ and $b$ and in-phase tilting along $c$.\cite%
{KHC09,ZKS10b} This explanation cannot be used for PbZrO$_{3}$, whose tilt
pattern $a^{-}a^{-}c^{0}$ in the O-AFE lacks the in-phase rotation along $c$%
, and whose parent phase is generally considered untilted.

Our explanation\cite{150} for stiffening in PZT is that in the R-FE phase
the octahedra are tilted about all three principal cubic directions, but
without long range order, so that the tilting is not easily observable by
diffraction. At present there is no direct indication of disordered tilting
in the $R3m$ phase of PZT with $x<0.06$, apart from controversial\cite{WKR05}
observations by TEM of in-phase tilting of the $a^{0}a^{0}c^{+}$ type,\cite%
{Vie95} and theoretical indications that the $R3m$ phase has strong tendency
to tilt.\cite{LCW02} Yet, the elastic compliance of PZT with $0.06\leq x\leq
0.15 $ clearly shows a transition at a temperature $T_{\mathrm{IT}}$, which
can only be associated with octahedral tilting.\cite{145,148} This
instability line is the continuation of $T_{\mathrm{T}}\left( x\right) $
when it starts dropping below $x\sim 0.15$ and merges with $T_{\mathrm{C}}$
at $x=0.06$. It follows that, when $x<0.06$, the FE transition at $T_{%
\mathrm{C}}$ also involves disordered tilting.\cite{148} Then, the OT
transition to the $a^{-}a^{-}c^{0}$ pattern occurs through the loss of
tilting about the $c$ axis, and may well cause stiffening.

\subsection{Fits of elastic anomalies}

In what follows we present several pairs of successive heating and cooling
cycles with the corresponding curves of the compliance and of the losses,
extending the range of shapes of the elastic anomalies with respect to the
previous work on PZT5.\cite{150} In all cases it is possible to fit the
curves with the hypothesis that the compliance $s$ is the average of the
compliances $s_{k}$ of the coexisting phases, weighted with their volume
fractions $f_{k}$:

\begin{equation*}
s=\sum_{k}f_{k}s_{k}
\end{equation*}%
with $k=$ FE (untilted R-FE $R3m$), T ($a^{-}a^{-}a^{-}$ tilting of the R-FE
$R3c$ phase), AFE (only cation displacements without long range $%
a^{-}a^{-}c^{0}$ tilt order) and OT (long range $a^{-}a^{-}c^{0}$ tilt
order). The reference compliance is $s_{\mathrm{FE}}$ of the $R3m$ phase,
assumed to include the anharmonic dependence on temperature; the latter is
taken as linear. Notice that some phases are mutually exclusive and some are
not: the FE and AFE cation orderings and the T ($a^{-}a^{-}a^{-}$) and OT ($%
a^{-}a^{-}c^{0}$) tilt patterns are exclusive pairs, but T may in principle
persist within AFE domains. These conditions are reflected in the volume
fractions. In some cases it is found necessary to introduce an interaction
between the AFD and polar orderings, with the assumption that the OT tilting
occurs only within AFE domains. If the volume fractions change sharply
between 0 and 1 at the transition temperatures, one gets steplike changes of
the elastic constant at the transitions, as expected from the Landau theory
for phase transitions with order parameters whose square is linearly coupled
to strain;\cite{Reh73,KHC09} this is indeed the case of the AFD and polar
transitions. The steplike elastic anomaly is combined with the condition
that each transition occurs over a certain time and temperature range,
setting the volume fractions $f_{k}$ as steps centered at the transition
temperatures $T_{k}$ with widths $\Delta T_{k}$,
\begin{equation}
f_{k}=\frac{1}{2}\left[ 1-\tanh \left( \frac{T-T_{k}}{\Delta T_{k}}\right) %
\right] ~.  \label{step}
\end{equation}%
The variety of the shapes of the anomalies derives from the strong
dependence of $T_{k}$ and $\Delta T_{k}$ on the sample state and temperature
rate. Certainly expressions (\ref{step}) is a simplification, especially
when describing transitions over broad temperature ranges and measured at a
variable temperature rate, but it will be shown that even with this
approximation remarkably good fits are possible. When the OT pattern sets
in, it replaces the T pattern, and therefore $f_{\mathrm{T}}$ has to be
multiplied by $1-f_{\mathrm{OT}}$. The fitting formula becomes
\begin{equation*}
s=b_{0}+b_{1}T+f_{\mathrm{T}}\left( 1-f_{\mathrm{OT}}\right) A_{\mathrm{T}%
}+f_{\mathrm{OT}}A_{\mathrm{OT}}+f_{\mathrm{AF}}A_{\mathrm{AF}}
\end{equation*}%
where $A_{k}=s_{k}-s_{\mathrm{FE}}$ are the amplitudes of the steps. In
certain cases it appears that the OT pattern does not sets in independently
of the AFE or FE order, but only after the occurrence of the AFE transition.
In these cases the OT contribution is set as $f_{\mathrm{OT}}f_{\mathrm{AF}%
}A_{\mathrm{OT}}$. In many cases it is clear that the transition to the AFE
or OT state is incomplete during cooling, and in these cases the
corresponding volume fraction is let to increase up to $f_{\mathrm{AF}}<1$.

In the following series of figures, the temperature dependences of the
compliance and of the losses are presented in pairs of successive heating
and cooling cycles with rates of $\sim 1$~K/min unless otherwise specified.
The continuous lines are fits to the real parts, while no attempt has been
done to fit the losses. The latter are in some case noisy and therefore the $%
s^{\prime \prime }/s^{\prime }$ curves are splines passing through the data
points. The vertical arrows indicate the transition temperatures $T_{k}$,
while the error bars extend between $T_{k}\pm \Delta T_{k}$.

\begin{figure}[tbh]
\includegraphics[width=8.5 cm]{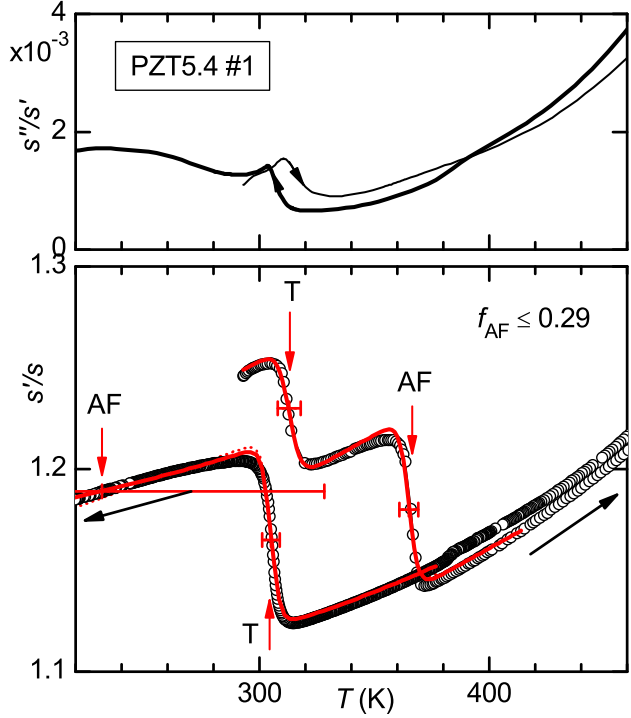}
\caption{(Color online) Elastic compliance of PZT5.4 \#1 during a
heating/cooling cycle in the virgin state. The continuous lines of $%
s^{\prime \prime }/s^{\prime }$ are splines through the experimental points;
the experimental points of $s^{\prime }/s_{0}$ (curves 1 and 2 of Fig.
\protect\ref{ti5.4n1}) are fitted with the continuous lines. The vertical
arrows indicate the fitting values $T_{k}$ of the steps and the error bars
their widths $\pm \Delta T_{k}$; the fraction of AFE phase was limited to $%
f_{\mathrm{AF}}\leq 0.29$ instead of 1. Heating and cooling rates of
approximately 1~K/min.}
\label{plotP34}
\end{figure}

We start with the first heat/cool cycle on sample PZT5.4 \#1 (Fig. \ref%
{plotP34} and curves 1 and 2 of Fig. \ref{ti5.4n1},). In this initial state
a partial transformation to the AFE state had occurred, and hence both the
stiffening at $T_{\mathrm{T}}^{h}=313$~K from the untransformed R-FE phase
and the stiffening at $T_{\mathrm{AF}}^{h}=365$~K are present. The ramp was
extended up to 600~K (well above $T_{\mathrm{C}}=513$~K), therefore erasing
the memory of the AFE domains, and no evidence is found of AFE or OT
transformation during the subsequent cooling. The initial AFE fraction $f_{%
\mathrm{AF}}=0.29$ can be deduced from the ratio of the initial step at $T_{%
\mathrm{T}}^{h}$ to that at $T_{\mathrm{T}}^{c}$, which is equal to $1-f_{%
\mathrm{AF}}$. A slight improvement of the fit for the cooling curve is
obtained assuming a broad AF transition centered at $T_{\mathrm{AF}%
}^{c}\simeq 230$~K, whose high temperature tail improves the shape of the
step.

\begin{figure}[tbh]
\includegraphics[width=8.5 cm]{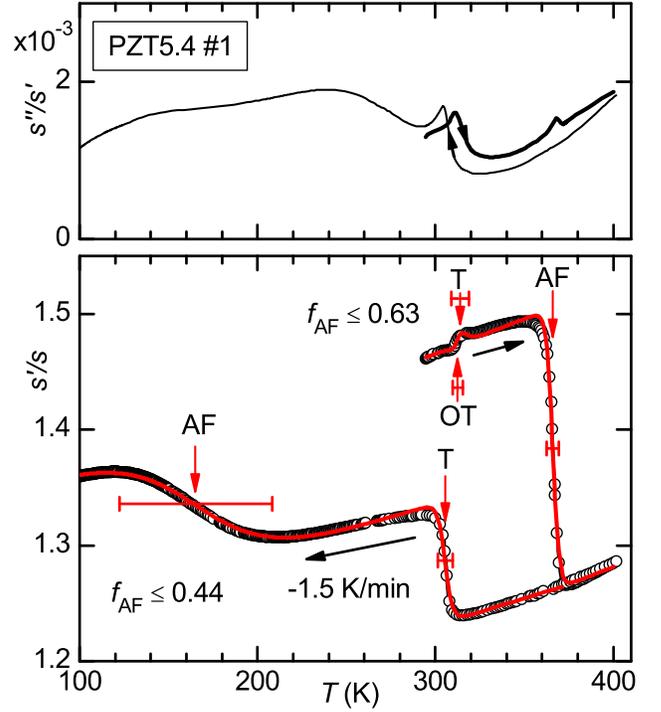}
\caption{(Color online) Elastic compliance of PZT 5.4 \#1 measured 47 days
after Fig. \protect\ref{plotP34}.}
\label{plotP45}
\end{figure}

\begin{figure}[hbt]
\includegraphics[width=8.5 cm]{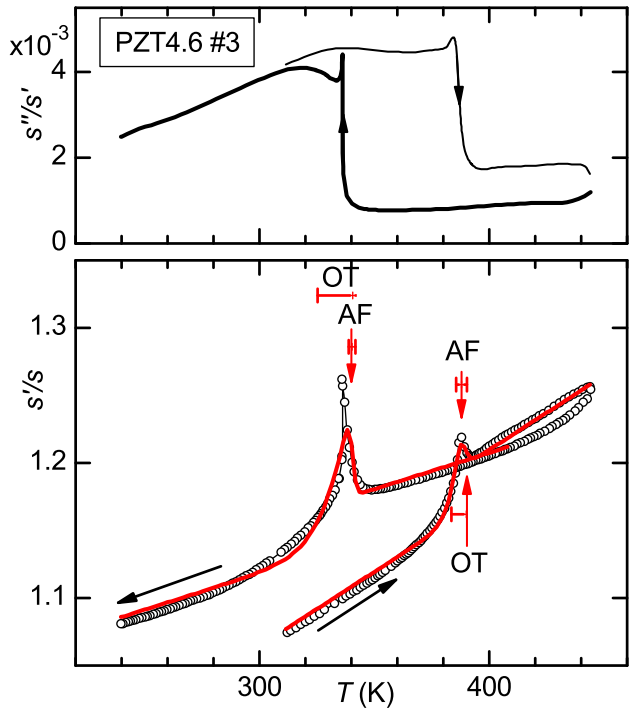}
\caption{(Color online) Anelastic spectrum of PZT4.6 \#3 5 days after
annealing up to 860~K. }
\label{plotP58}
\end{figure}

The next curves (Fig. \ref{plotP45} corresponding to curves 3 and 4 of Fig. %
\ref{ti5.4n1}) are measured after 47 days, during which almost complete
transformation to the AF+OT phase should have occurred, judging from the
seeming lack of the negative step at $T_{\mathrm{T}}^{h}=313$~K (see \textit{%
e.g.} Fig. \ref{plotP34}). In its place, a small positive step with a
wavelet is found, which suggests the mutual cancelation of OT and T steps at
nearly the same temperature but with different widths. In most cases, the OT
and AF transitions occur together during heating (Ref. \onlinecite{150},
curve \#1 in Fig. \ref{ti5.4n1n3}, Figs. \ref{plotP41} and \ref{plotP58}),
but in sample PZT5.4 \#1 the OT tilting appears less stable, and it seems
that in Fig. \ref{plotP45} the disappearance of $a^{-}a^{-}c^{0}$ tilting is
triggered by the disappearance of $a^{-}a^{-}a^{-}$ tilting in the
neighboring R-FE domains at $T_{\mathrm{T}}^{h}$, before the transition to
the FE state occurs. If one assumes that the OT component disappears
completely at $T_{\mathrm{OT}}^{h}\simeq T_{\mathrm{T}}^{h}$, there is
uncertainty on the magnitude of the mutually canceled T and OT steps, which
reflects in an uncertainty in the fraction $f_{\mathrm{AF}}$ of AFE phase in
the initial state. The fit of Fig. \ref{plotP45} is done setting $A_{\mathrm{%
OT}}=$ 0.1, similarly to other cases, which results in $f_{\mathrm{AF}%
}\simeq 0.63$ during heating and 0.44 during cooling, but equivalent fits
can be obtained by setting a smaller $A_{\mathrm{OT}}$, with the initial $f_{%
\mathrm{AF}}$ up to 1. During cooling no sign of transformation to the OT
phase is found. As before, the losses present a steplike increase in the
tilted R phase, $T<T_{\mathrm{T}}$, and an anomaly is hardly visible at $T_{%
\mathrm{AF}}^{h}$.

\begin{figure}[tbh]
\includegraphics[width=8.5 cm]{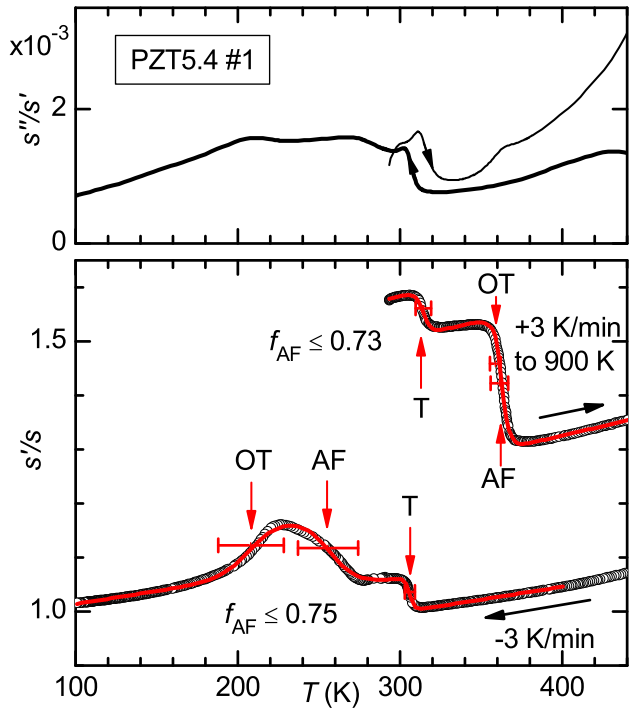}
\caption{(Color online) Curves 5 and 6 of Fig. \protect\ref{ti5.4n1},
including heating up to 900~K.}
\label{plotP47}
\end{figure}

Figure \ref{plotP47} corresponds to curves 5 and 6 of Fig. \ref{ti5.4n1},
measured at $\pm 3$~K/min; 6 days after the previous run, a considerable
fraction of AF+OT phase had formed, but there is ambiguity on how much the
two transitions contribute to the step at $T_{\mathrm{AF}}^{h}=$ $361.3$~K$%
~\simeq T_{\mathrm{OT}}^{h}$. In fact, the heating was extended to 900~K,
causing a recover of the overall compliance and of the compliance steps to
the original lower values. Therefore, it was not possible to share in the
fit the amplitudes of the steps between heating and cooling, as in all other
cases. We chose to share only $A_{\mathrm{OT}}$. Notice that, thanks to the
dissolution at 900~K of the defects that stabilized the FE\ phase, both the
AF and OT transitions are observed even at a cooling rate as fast as
3~K/min. It is possible to obtain very similar fits of the broad split AF+OT
transition both with and without the condition that the OT is subordinated
to AF; the only parameters that are affected by the choice are the step
amplitudes $A_{\mathrm{AF}}$ and $A_{\mathrm{OT}}$, which change by less
than 10\%.

\begin{figure}[tbh]
\includegraphics[width=8.5 cm]{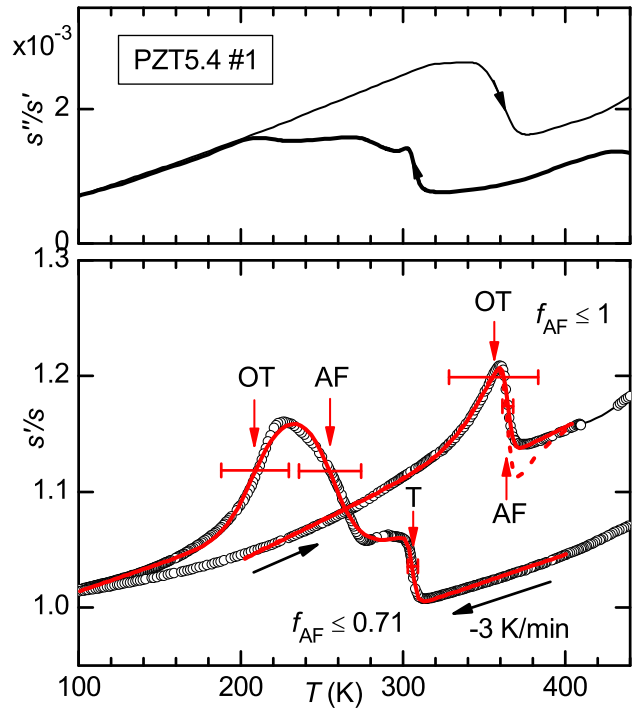}
\caption{(Color online) Curves 6 and 7 of Fig. \protect\ref{ti5.4n1}. The
dashed line is obtained assuming that AF and OT are independent of each
other.}
\label{plotP48}
\end{figure}

The final fit of the PZT5.4 \#1 sample (Fig. \ref{plotP48}) is a combination
of the previous cooling in the annealed state and the subsequent heating.
Also in this case it was impossible to share the amplitude $A_{\mathrm{AF}}$
between cooling and heating, because its magnitude is clearly greater during
cooling. In this case the heating curve clearly indicates that the OT order
is subordinated to the AF order. In fact, the slow upturn during heating is
fitted with a broad OT anomaly, centered at a temperature $T_{\mathrm{OT}%
}^{h}$ very close to $T_{\mathrm{AF}}^{h}$, as usually is the case during
heating, but the slow upturn would continue also above $T_{\mathrm{AF}}^{h}$
as in the dashed curve. The maximum fraction of transformed AFE phase on
cooling was not imposed and left as a free parameter, but its value 0.71 is
consistent with that of the previous fit ($f_{\mathrm{AF}}=0.76$). These
curves demonstrate that some interaction between the polar and AFD modes
exists, and it is not always possible to fit elastic anomalies overlapping
over several tens of kelvin as if they were completely independent of each
other.

\begin{figure}[hbt]
\includegraphics[width=8.5 cm]{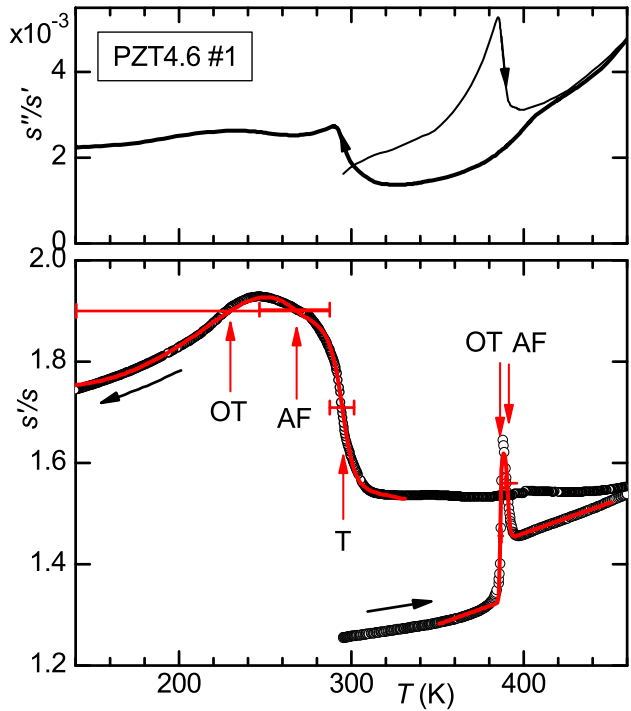}
\caption{(Color online) Anelastic spectrum of PZT4.6 \#1 during a
heating/cooling cycle in is virgin state.}
\label{P37}
\end{figure}

We pass with Fig. \ref{P37} to PZT4.6 \#1 in its virgin state. At this
composition $T_{\mathrm{T}}\simeq T_{\mathrm{AF}}^{c,0}\simeq $ room
temperature, but from the lack of stiffening at $T_{\mathrm{T}}^{h}$ it is
deduced that the sample was initially completely transformed in the AF+OT
state and reverted sharply to the untilted FE phase. The combination of
stiffening (AF$\rightarrow $FE) and softening (loss of long range OT tilt
order) at practically the same temperature produces a spike in the
compliance. The heating was extended to $T_{\mathrm{C}}+50$~K~= 560~K, but
contrary to the PZT5.4 and also PZT5 cases, both the AF and OT transitions
were complete, though broadened, also during cooling. All the previous fits
to the PZT5.4 curves were made imposing that OT tilting takes place only
within AF domains (see especially Fig. \ref{plotP48}), but the fit to the
virgin PZT4.6 \#1 is much better by relaxing this condition. This was also
the case of PZT5, where sharp OT ordering occurred without observable
formation of AF order.

\begin{figure}[hbt]
\includegraphics[width=8.5 cm]{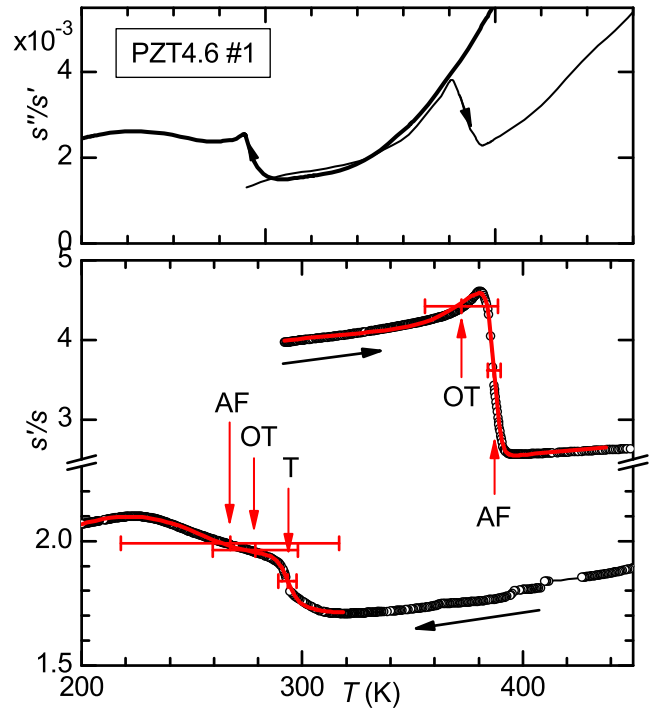}
\caption{(Color online) Anelastic spectrum of PZT4.6 \#1 25 days after the
measurement of Fig. \protect\ref{P37}. The cycle was extended to 900~K}
\label{plotP41}
\end{figure}

After 25 days of aging the compliance at room temperature had increased of
more than three times (Fig. \ref{plotP41}), which is really an enormous
effect, with a softening mainly developed in the AF phase, but also of the
FE phase became 1.7 times softer than initially. The run was extended to
625~K and produced a partial anneal of the defects and recovery of the
stiffness. During cooling, the AF and OT transitions were visible but
broadened and possibly only partial: the fit was done by sharing the
amplitudes of the steps but reducing the fraction $f_{\mathrm{AF}}\leq 0.55$
during cooling, but it is also possible that the amplitudes of the steps
were halved by the partial annealing and the fraction of transformed phase
was $f_{\mathrm{AF}}\simeq 1$. As discussed in reference to Fig. \ref%
{plotP48}, the broad OT transition on heating had to be forced to follow the
sharper AF one, in order to avoid an undershoot at $T>T_{\mathrm{AF}}^{h}$.

The last pair of fits (Fig. \ref{plotP58}) is from another sample, PZT4.6
\#3, measured after high-temperature annealing. The amplitudes of the AF and
OT steps are very close to those of the virgin PZT4.6 \#1, the main
difference being that virgin PZT4.6 \#1 had broad transitions during
cooling, while annealed PZT4.6 \#3 has a cooling curve similar to the
heating one, with sharp and almost coincident AF and OT transitions.

\section{Conclusions}

The phase diagram of PZT near the boundary between FE and AFE phases is the
result of the action of polar, antipolar and octahedral tilt instabilities,
but also by yet scarcely characterized defects. Such defects, which we
tentatively identify with variously aggregated O vacancies, can drastically
change the kinetics of the different modes involved in the transition to the
AFE state, causing a freezing of the tilt or antipolar modes or both, but
also more than doubling in a reversible way the elastic compliance. This is
in marked contrast with the aging usually found in glassy and disordered
systems, but also FE or relaxor PZT,\cite{115,122} which is accompanied by a
decrease of the susceptibility or stiffening.

As a result of the interplay of the different polar and tilt modes and
probably defect aggregation, the curves of the elastic compliance versus
temperature exhibit a variety of shapes, which however can be well fitted
with simple hypotheses for the studied compositions $0.046\leq $ $x\leq
0.054 $. There are three transitions connected with the establishment of 1)
polar/antipolar order, 2) octahedral tilt pattern in the AFE orthorhombic
phase ($a^{-}a^{-}c^{0}$, here labeled OT), and 3) tilt pattern in the FE
rhombohedral phase ($a^{-}a^{-}a^{-}$). Each transition causes a steplike
change in the elastic compliance, whose temperature and width depend on the
kinetics of the transition. These steps can be well approximated with
hyperbolic tangents of $T$, whether sharp or broad, and the imperfect fits
are probably due to variations in the temperature rates as much as to the
choice of the hyperbolic tangent as fitting function.

In most of the presently studied samples the octahedral tilt order OT seems
less stable and subordinated to the AFE one, but there are cases where the
fit requires that the two are independent (Fig. \ref{P37}), and this is
especially true in the previous study on PZT5,\cite{150} where sharp OT
transitions occur in the absence of the AFE ones. A case is also presented
(Fig. \ref{plotP45}) where the loss of OT tilting appears to be triggered by
the loss of tilting in the coexisting R-FE domains.

The slow kinetics for the onset of the AF+OT order is due to some type of
aggregation of mobile defects, which causes a stabilization of the FE phase
and a general softening of the lattice, by factors as large as 2.3 within
the FE phase and even 4 at room temperature, where in addition the state can
vary between FE and AFE. We emphasize that such macroscopic changes of the
elastic properties occur simply during room temperature aging or after few
FE/AFE cycles, and not after repeated electric field induced cycling as in
fatigue experiments. Moderate annealing at 900~K, well below the sintering
temperature, can recover the pristine elastic stiffness and promptness to
form the AFE phase. This fact suggests that the defects involved are O
vacancies rather than cracks, chemical decomposition or cation diffusion,
which instead are relevant in fatigue experiments.

\begin{acknowledgments}
The authors thank C. Capiani (ISTEC) for the skillful preparation of the
samples, P.M. Latino (ISC) and R. Scaccia (ISC) for their technical
assistance in the anelastic and dielectric experiments.
\end{acknowledgments}

\end{document}